# Laser-induced electron diffraction of the ultrafast umbrella motion in ammonia

B. Belsa[1], K. Amini[1,2], X. Liu[1], A. Sanchez[1] T. Steinle[1], J. Steinmetzer[3], A.T. Le[4], R. Moshammer[5], T. Pfeifer[5], J. Ullrich[5], R. Moszynski[2], C.D. Lin[6], S. Gräfe[3], J. Biegert[1,7,†]

[1]ICFO - Institut de Ciencies Fotoniques, The Barcelona Institute of Science and Technology, 08860 Castelldefels (Barcelona), Spain.
[2]Department of Chemistry, University of Warsaw, 02-093 Warsaw, Poland.
[3]Institute of Physical Chemistry and Abbe Center of Photonics, Friedrich-Schiller-Universität Jena, Helmholtzweg 4, 07743 Jena, Germany.
[4]Department of Physics, Missouri University of Science and Technology, Rolla, MO 65409.
[5]Max-Planck-Institut für Kernphysik, Saupfercheckweg 1, 69117, Heidelberg, Germany.
[6]Department of Physics, J. R. Macdonald Laboratory, Kansas State University, 66506-2604 Manhattan, KS, USA.
[7]ICREA, Pg. Lluís Companys 23, 08010 Barcelona, Spain.

[†]To whom correspondence should be addressed to. Email: jens.biegert@icfo.eu.

## ABSTRACT

Visualizing molecular transformations in real-time requires a structural retrieval method with Ångström spatial and femtosecond temporal atomic resolution. Imaging of hydrogen-containing molecules additionally requires an imaging method that is sensitive to the atomic positions of hydrogen nuclei, with most methods possessing relatively low sensitivity to hydrogen scattering. Laser-induced electron diffraction (LIED) is a table-top technique that can image ultrafast structural changes of gas-phase polyatomic molecules with sub-Ångström and femtosecond spatiotemporal resolution together with relatively high sensitivity to hydrogen scattering. Here, we image the umbrella motion of an isolated ammonia molecule ($NH_3$) following its strong-field ionization. Upon ionization of a neutral ammonia molecule, the ammonia cation ($NH_3^+$) undergoes an ultrafast geometrical transformation from a pyramidal ($\Phi_{HNH} = 107°$) to planar ($\Phi_{HNH} = 120°$) structure in approximately 8 femtoseconds. Using LIED, we retrieve a near-planar ($\Phi_{HNH} = 117 \pm 5°$) field-dressed $NH_3^+$ molecular structure 7.8 – 9.8 femtoseconds after ionization. Our measured field-dressed $NH_3^+$ structure is in excellent agreement with our calculated equilibrium field-dressed structure using quantum chemical *ab initio* calculations.

## I. INTRODUCTION

Many important processes in nature rely on the motion of hydrogen atoms, such as the influence of proton dynamics on the biological function of proteins[1,2]. The motion of the hydrogen atom, which is the lightest element in the periodic table, occurs on the few-femtosecond (few-fs; 1 fs = $10^{-15}$ s) timescale, and represents the fastest possible nuclear motion in molecules. Consequently, a method is required that is both sensitive and fast enough to probe the motion of hydrogen atoms with sub-Ångstrom (sub-Å; 1 Å = $10^{-10}$ m) spatial and femtosecond temporal atomic resolutions[3-4]. The static geometric structure of molecules can be successfully determined through a variety of imaging and spectroscopic techniques[5], such as conventional electron diffraction (CED)[6], X-ray diffraction and crystallography[7], optical and nuclear magnetic resonance (NMR) spectroscopies[8], scanning tunneling microscopy (STM)[8] and atomic force microscopy (AFM)[8]. In particular, the time-resolved analogues of X-ray and electron diffraction, such as ultrafast X-ray diffraction (UXD)[9,10] and ultrafast electron diffraction (UED)[11–18], have provided a wealth of dynamical information in molecules that contain atoms much heavier than hydrogen. As a result, their scattering signal in such molecules is very large and their respective dynamics occur on the hundreds-of-femtosecond timescale.

Laser-induced electron diffraction (LIED)[5,19–30] is a strong-field variant of UED that can directly retrieve the geometric structure of gas-phase molecules containing hydrogen atoms with sub-Å and few-to-sub-fs spatiotemporal resolution. The LIED technique is based on probing the molecular geometric structure using the molecule's own emitted electron to elastically scatter against the atomic cores in the molecule during strong-field-induced recollisions. The intra-optical-cycle nature of the LIED process enables structural retrieval with sub-femtosecond time resolution. Moreover, because of the small de Broglie wavelength of the electrons, the technique provides picometer (pm; 1 pm = $10^{-12}$ m) spatial resolution. Importantly, LIED is sensitive to hydrogen atom scattering as the kinetic energy of scattering electrons in LIED (*i.e.* 50 – 500 eV) are significantly lower than the tens or hundreds of keV used in UED. At these low impact energies, hydrogen exhibits significant scattering cross-section values compared to those at the high energies and forward-only scattering employed in UED. Presently, improving sensitivity to hydrogen scattering with other methods is challenging. Moreover, the low-energy nature of LIED electrons also provides a probe of the angular dependence of elastic electron scattering and thus the extraction of doubly-differential scattering cross-sections.

In the present work, we demonstrate LIED's capability to image the motion of hydrogen atoms on the few-fs timescale by studying the umbrella (inversion) motion of the ammonia molecule ($NH_3$) following its strong-field ionization. Photoelectron spectra and photoionization-induced dynamics of individual ammonia molecules and clusters have been a topic of interest in the past decades, both experimentally[31–35] and theoretically[34–40]. Neutral ammonia at its equilibrium configuration has a pyramidal shape, described by the $C_{3v}$ symmetry point group, with an equilibrium H-N-H bond angle[41], $\Phi_{HNH}$, of 107°, as shown in Fig. 1. When ionized, the ammonia molecule undergoes a significant geometrical transformation as the ammonia cation in its ground electronic state has a planar equilibrium geometry of $D_{3h}$ symmetry with an equilibrium $\Phi_{HNH}$ of 120°. Förster and Saenz (2013) developed a theoretical model to describe the inversion motion of the ammonia cation ($NH_3^+$) for high-harmonic spectroscopy where they predict that the $NH_3^+$ nuclear wave packet reaches the potential minimum on a 5-fs timescale[36]. Kraus and Wörner (2013) theoretically investigated the pyramidal-to-planar transition in $NH_3^+$, which they calculated to occur on a 7.9-fs timescale. The authors also experimentally studied the same dynamics but could only indirectly provide partial evidence of the umbrella motion through high-harmonic spectroscopy (HHS). These HHS measurements were in fact performed at different wavelengths in the near-infrared (NIR) up to 1.8 μm, reaching a temporal range of up to 3.8 fs after ionization to be investigated[34]. In the aforementioned HHS studies, structural information could only be indirectly obtained, with no direct imaging studies previously reported. Here, we use MIR-LIED to directly retrieve structural information of the $NH_3^+$ cation 7.8 – 9.8 fs after ionization, which is on a similar (7.9-fs) timescale as that predicted for the pyramidal-to-planar transition to occur in the $NH_3^+$ cation[34].

This paper is organized as follows: first, a brief overview of the experimental setup and the theoretical methods employed in this work is given in Section II, followed by a discussion of the experimental and theoretical results in Section III, and finally, a summary and conclusion of our results are presented in Section IV.

## II. EXPERIMENTAL AND COMPUTATIONAL METHODS

### A. Mid-infrared (MIR) OPCPA source

The MIR laser source is a home-built optical parametric chirped-pulse amplifier (OPCPA) that has been previously described[42]. Briefly, the OPCPA set-up generates a 3.2 μm laser pulse with a duration of 100 fs full width at half maximum (FWHM) at a 160 kHz repetition rate. The high repetition rate compensates for the reduced rescattering cross-section due to the $\lambda^{-4}$ scaling factor[43]. The laser pulse is focused into the molecular beam using an on-axis paraboloid that is placed inside of the reaction microscope. The focal spot size achieved was 6 – 7 μm, resulting in a peak intensity, $I_0$, of 1.3 x 10$^{14}$ W/cm$^2$. Such peak intensity translates to a ponderomotive energy (*i.e.* the average kinetic energy of a free electron in an oscillating electric field), $U_p$, of 120 eV, which corresponds to the maximum classical return energy ($E_r^{max} = 3.17 U_p$) of about 380 eV, and the maximum backscattered energy ($E_{resc}^{max} = 10 U_p$) of 1200 eV. The Keldysh parameter[44], $\gamma = \sqrt{I_p/(2U_p)}$, was approximately 0.2.

### B. Reaction Microscope detection system

The detection system is based on a reaction microscope (ReMi)[45], which has been previously described in detail elsewhere[46] with only a brief summary presented here. A cold ($T < 100K$) ammonia jet (5% NH$_3$, 95% He) was supersonically expanded into an ultra-high vacuum (UHV) chamber. Here, the interaction with the laser focus takes place, ionizing the gas. Upon strong-field ionization, the generated ions and electrons were guided using homogeneous electric ($\vec{E}$) and magnetic ($\vec{B}$) fields of 34 V/cm and 13 G, respectively, towards two opposing time-sensitive microchannel plates (MCPs) sensors. These sensors are interfaced with position-sensitive delay-line anode detectors. The three-dimensional (3D) momentum distribution $(p_x, p_y, p_z)$ of charged particles is extracted from the time-of-flight (ToF; parallel to the $z$-axis) and the $(x, y)$ impact position on the two-dimensional (2D) detector plane. Charged particles are detected in full electron-ion coincidence, enabling the isolation of different reaction paths.

### C. Theoretical framework of LIED retrieval

LIED is a strong field technique in which a rescattering electron acquires structural information when scattered off its target in the presence of a laser field. Therefore, measured momenta contain two contributions, a momentum shift due to scattering off the target molecule, and a momentum shift due to the vector potential of the laser at the time of rescattering. The value of the vector potential varies during the laser cycle and thus imparts different momentum at the varying times of rescattering during the laser cycle. Under quasi-static (tunneling) conditions, the exact time variation can however be determined with very good accuracy from the classical recollision model. The vector potential can be extracted from a measurement of the laser's peak intensity directly. We employ another, more accurate way, to extract the vector potential directly from the identification of the $2U_p$ turnover between direct electrons and the rescattering plateau, and the cutoff at $10U_p$. The value of the electric field, and thus the vector potential, is found according to $U_p = E_0/4\omega^2$. In addition, we determine the laser peak intensity by fitting the momentum dependent ionization rate calculated after the ADK theory to the longitudinal ion momentum distribution $(p_\parallel)$ of Ar$^+$ ions. Both methods are work very well and yield $I_0 = 1.3 \times 10^{14}\ W/cm^2$ ($E_0 = 0.06\ a.u.$). As already mentioned, the measured electron rescatters from the parent ion in the laser field where it receives an additional momentum kick from the laser related to its vector potential, $A_r(t_r)$, at the time of rescattering, $t_r$, in polarization direction. Therefore, the final detected momentum, $k_\parallel$ ($k_\perp$), parallel (perpendicular) to the laser polarization direction is related to the return momentum, $k_r$, and scattering angle, $\theta_r$, as $k_\parallel = -A_r(t_r) \pm k_r \cos(\theta_r)$ and $k_\perp = k_r \sin(\theta_r)$. FT-LIED is based on the measurement of backscattered electrons (i.e., for $\theta_r = 180°$), thus yields $k_r = k_{resc} - A_r(t_r)$, where $A_r(t_r)$ is calculated for a detected momentum, $k_{resc}$, employing the classical recollision model which is valid under our quasi-static field conditions. The position $x(t, t_b)$ of the electron in a linearly polarized electric field can be obtained from the classical equation of motion according to $x(t, t_b) = \frac{E_0}{\omega^2}[\sin(\omega t_b)(\omega \Delta t) + \cos(\omega t) - \cos(\omega t_b)] + v_0 \Delta t + x_0$, where $t_b$ and $t$ are the time of birth and time in the laser field, respectively, and $\Delta t$ is the difference between $t_b$ and $t$. For quasi-static conditions, the initial velocity, $v_0$, of the electron at the tunnel exit, $x_0$, is assumed zero. An electron will return to the parent ion when $x(t_r, t_b) = 0$ at the time of rescattering, $t_r$. The equation of motion can be solved numerically by Newton's method and general solutions are found for electrons tunnelling between $0 \leq t_b \leq 0.25$ of an optical cycle and returning between $0.25 \leq t_r \leq 1$ of the optical cycle. For a given $t_b$, its corresponding $t_r$ is calculated. In general, there exist two trajectories, called long and short, which lead to the same final momenta. However, the long trajectory is born much closer to the maximum of the laser field. The exponentially dependent ionization yield thus favours the early ionizing long trajectory which is the reason why the short trajectory contribution is neglected. With one trajectory present, the vector potential at $t_r$ is obtained as $A_r = -\frac{E_0}{\omega}\sin(\omega t_r)$, making straightforward the reconstruction of $k_r$ and allowing to unambiguously map momentum to time of rescattering.

### D. Quantum chemistry calculations

The adiabatic ground state potential energy surfaces (PESs) along the inversion coordinate of both neutral NH$_3$ ($\tilde{X}\,^1A_1'$) and cation NH$_3^+$ ($\tilde{X}\,^2A_2''$) were calculated at the coupled cluster singles doubles (CCSD)[47] level of theory as implemented in the Q-Chem 5.1 quantum chemistry package[48]. The augmented correlation-consistent, polarized valence, double-zeta Dunning basis set (aug-cc-pVDZ)[49] was applied. Permanent dipole moments ($\mu_x$, $\mu_y$, $\mu_z$), as well as static dipole polarizabilities ($\alpha_{xx}$, $\alpha_{yy}$, $\alpha_{zz}$), were calculated at all points of the potential energy surface (PES). The field-dressed energies were calculated as follows

$$E(F) = E_0 - \mu \cdot F - \frac{1}{2}\alpha \cdot F^2 \quad [1]$$

where $F$ is the electric field strength, $E_0$ is the field-free Born-Oppenheimer energy, $\mu$ the permanent dipole moment and $\alpha$ the main diagonal of the polarizability tensor. The field strength was set to 0.06 a.u. (3.1 V/Å), corresponding to a laser peak intensity of 1.3 x 10$^{14}$ W/cm$^2$.

All geometries were previously optimized at the second-order Møller-Plesset (MP2) level of theory using the ANO-RCC-VDZP[50] basis set in OpenMolcas 8.0[51]. A dummy atom (X) was placed along the $z$-axis, which is parallel to the $C_3$ principal axis of NH$_3$, at a distance of 1.0 Å above the nitrogen atom (N). The H1-N-X, H2-N-X and H3-N-X angles ($\beta$) were constrained to vary from 130° to 90° in steps of 1° (*i.e.* total of 41 geometries). Planarity is therefore defined by $\beta = 90°$. Here, the inversion coordinate, $Q$, is defined as displacement with respect to the reference geometry in degrees, where a value of 0° corresponds to planarity (*i.e.* $\Phi_{HNH} = 120°$). For a negative (positive) value of $Q$, the nitrogen atom is located above (below) the plane spanned by the three hydrogen atoms. A sketch of the described coordinate is shown in Fig. 5.

## III. RESULTS AND DISCUSSION

### IIIa. FT-LIED analysis

The procedure for retrieving structural information is based on the Fourier transform (FT) variant of LIED, called FT-LIED[20,23] which is also known as the fixed-angle broadband laser-driven electron scattering (FABLES)[25] method. In the FT-LIED method, Fourier transforming the coherent molecular interference signal, $\rho_M$, embedded within the momentum distribution of the backscattered highly-energetic electrons (i.e. $\theta_r = 180°$) directly provides an image of the molecular structure in the far-field. The key benefit of the FT-LIED scheme is its ability to empirically retrieve the background incoherent sum of atomic scatterings, $\rho_A$, that contributes to the total detected interference signal, $\rho_E$. Thus, the $\rho_M$ can be obtained from $\rho_E$ by subtracting the empirically retrieved $\rho_A$ to directly retrieve the molecular structure without the use of theoretical fitting, retrieval or modelling algorithms.

Fig. 2 shows the logarithmically scaled momentum distribution of longitudinal ($P_\parallel$; parallel to the laser polarization) and transverse ($P_\perp$; perpendicular to the laser polarization) momenta for electrons detected in coincidence with the $NH_3^+$ molecular ion. *Direct* electrons oscillate away from the parent ion without rescattering, with a momentum obtained initially by the vector potential of the laser field, $A(t_r)$, at the instance of rescattering, $t_r$. Hence, the maximum kinetic energy that the electron can gain is $2U_p$ (*i.e.* momentum $P_\parallel \leq 4.2$ a.u). *Rescattered* electrons, however, propagate further in the field, acquiring significantly higher kinetic energy after recolliding against the parent ion at $t_r$ with an appreciably large return momentum, $k_r$. Additionally, the rescattered electron is also "kicked" by the laser field at $t_r$, receiving an additional momentum in the polarization direction. Therefore, the total detected momentum, $k_{\text{resc}}$, is related to the return momentum at the instance of rescattering, $k_r$, and the momentum "kick" obtained by the vector potential, $A(t_r)$, of the laser field through $k_{\text{resc}} = k_r + A(t_r)$ (see the sketch in Fig. 2). The overall maximum kinetic energy obtained by the *rescattered* electrons is ten times the ponderomotive potential ($10U_p$) (*i.e.* $P_\parallel = 9.4$ a.u.) for *backscattered* electrons. In this sense, the elastically rescattered electrons, which contain structural information, can be distinguished from the direct electrons in the kinetic energy spectrum (momentum distribution) for energies of $2U_p \leq E_{resc} \leq 10U_p$ ($4.2 \leq P_\parallel \leq 9.4$ a.u.).

Since the FT-LIED method is applied, only coincidence electrons with a returning momentum of $k_r > 2.1$ a.u. (*i.e.* $P_\parallel > 4.2$ a.u.) and increasing rescattering angles, $\Delta\theta_r$, from 2 to 10° around the backscattering angle of $\theta_r = 180°$ are analyzed. At low $k_r$, a small $\Delta\theta_r$ is taken to avoid appreciable contributions from direct electrons which do not contain structural information. While at appreciably large enough $P_\parallel$, larger values of $\Delta\theta$ can be taken for higher $k_r$ since direct electrons do not contribute in this momentum region as they are significantly less energetic than the rescattered electrons. The interference signal is extracted by integrating an area indicated by a block arc in momentum space, as shown schematically in Fig. 2b, at various vector potential kicks, $A_r$.

### IIIb. Electron-ion 3D coincidence detection

In strong-field LIED studies, other events aside from elastic scattering of the tunnel-ionized electrons will occur. For example, more than one electron can be removed from the molecule, leading to the Coulomb explosion of multiply-charged $NH_3^{n+}$ and the subsequent production of $NH_2^+$ and $H^+$ ions and corresponding electrons. Moreover, there may also be contributions to the overall signal from background molecules existing in the main chamber, generating ion species that are not of interest in this study (*e.g.* $H_2O^+$, $N_2^+$ or $O_2^+$). All of these background ions and their corresponding electrons are detected in our spectrometer along with our molecular ion of interest, $NH_3^+$, which is the main peak at approximately 4.1 µs in the ion time-of-flight (ToF) spectrum shown in Fig. 3a. Electrons corresponding to background ions contribute an unwanted background signal in the FT-LIED analysis process, impeding structure retrieval when averaging over all molecular ionization channels. Electron-ion coincidence detection is implemented to ensure that the LIED interference signal originates only from our ion of interest (*i.e.* $NH_3^+$). To highlight the importance of coincidence detection, the total electron signal for all ions (petrol blue) and those electrons detected in coincidence with $NH_3^+$ (orange) are shown in Fig. 3b. In both distributions, the $2U_p$ and $10U_p$ classical cut-offs are clearly visible (vertical dashed lines). An order-of-magnitude difference in the number of electron counts is observed in the rescattering frame. Furthermore, the inset panel in Fig. 3b emphasizes the more pronounced oscillations, arising from the molecular interference signal, observed in the $NH_3^+$ coincidence distribution (orange) as compared to the 'all electrons' distribution (petrol blue).

**IIIc. Molecular structure retrieval**

Electrons detected in coincidence with $NH_3^+$ ions are plotted in Fig. 4a as a function of return kinetic energy in the range of 40 – 350 eV corresponding to the rescattering plateau of $2U_p$ – $10U_p$ range. The experimentally measured molecular backscattered electron distributions ($\rho_E$) (orange solid trace) contains contributions from both the incoherent sum of atomic scatterings – which is independent of molecular structure and thus, serves as a background ($\rho_B$) signal – and the coherent molecular interference signal ($\rho_M$). We calculate the LIED interference signal by subtracting an empirically determined background (by fitting a third-order polynomial function) from the logarithm of $\rho_E$[20], given by

$$\rho_M = \log_{10}(\rho_E) - \log_{10}(\rho_B) = \log_{10}(\rho_E/\rho_B) \quad [2],$$

and is plotted in Fig. 4b as a function of momentum transfer, $q = 2k_r$, in the back-rescattered frame. Observed oscillations in the interference signal (orange solid trace) provide a unique, sensitive signature of the molecular structure, with the orange (gray) shaded regions

The Fast Fourier transform (FFT) spectrum generated from the molecular interference signal, embedded within the interference signal (Fig. 4b), is shown in Fig. 4c. Before transforming, a Kaiser window[52] ($\beta = 0$) and zero padding[53] are applied. The FFT spectrum (orange solid trace), individual Gaussian fits (gray dotted traces) and the sum of the two Gaussian fits (blue solid trace) are presented. The center position of the individual Gaussian fits of the two FFT peaks appear at 1.31 ± 0.03 Å and 2.24 ± 0.03 Å, as shown in the sketch of Fig. 4d. Table 1 shows the N-H internuclear distance reported for neutral $NH_3$ in the ground electronic state and $NH_3^+$ cation in the ground and first excited electronic state. Comparing our FFT spectrum shown in Fig. 4c to the data in Table 1, it is clear that the first FFT peak at 1.31 ± 0.03 Å corresponds to the N-H internuclear distance, $R_{NH}$, whilst the second FFT peak at 2.24 ± 0.03 Å corresponds to the H-H internuclear distance, $R_{HH}$. The FT-LIED measured internuclear distances correspond to an H-N-H bond angle, $\Phi_{HNH}$, of 117 ± 5° (see Supplementary Materials for further details). We note that the peaks above 3 Å could arise from clusters of ammonia, which have been reported to have a centre-of-mass (i.e. N-N) distance of between 3.2-5.2 Å and full H-H distances within a cluster of up to 8 Å[54,55].

**Table 1 | Field-free equilibrium geometrical parameters of NH₃ and NH₃⁺.** The N-H and H-H internuclear distances, $R_{NH}$ and $R_{HH}$, respectively, and the H-N-H angle, $\Phi_{HNH}$, for neutral NH₃ in the ground electronic state[56]. The same geometric parameters for NH₃⁺ in the ground[57] electronic state is also presented.

|  | $R_{NH}(\text{Å})$ | $R_{HH}(\text{Å})$ | $\Phi_{HNH}(°)$ |
|---|---|---|---|
| NH₃ ($\widetilde{X}\,^1A_1'$) | 1.030 | 1.662 | 107 |
| NH₃⁺ ($\widetilde{X}\,^2A_2''$) | 1.023 | – | 120 |

### IIIe. Quantum chemistry calculations

To aid in our interpretation and understanding of the FT-LIED measured NH₃⁺ structure, we investigate the pyramidal-to-planar geometrical transition that ammonia undergoes following strong-field ionization. We perform quantum chemical *ab initio* calculations of field-free (black solid curves) and field-dressed (colored dashed curves) ground state potential energy curves (PECs) of neutral NH₃ (bottom panel) and NH₃⁺ cation (top panel), as shown in Fig. 5. The inversion coordinate ($Q$) employed is also shown at the bottom left side of Fig. 5.

A value of 0° corresponds to planarity. For $Q < 0$ ($Q > 0$), the nitrogen atom is located above (below) the plane spanned by the three hydrogen atoms. The pyramidal-to-planar transition is initiated at the time of ionization ($t = 0$ fs) where a nuclear wave packet (NWP) in the neutral ammonia is transferred to the PES of the NH₃⁺ cation. Kraus and Wörner (2013)[34] calculated that the NWP in field-free NH₃⁺ reaches the equilibrium planar structure (*i.e.* $Q = 0°$; $\Phi_{HNH} = 120°$) at 7.9 fs[34]. Thus, the equilibrium planar structure could be directly retrieved with MIR-LIED since the emitted LIED electron takes 7.8-9.8 fs to be accelerated and driven back to the NH₃⁺ parent ion by the laser field. In fact, we resolve an H-N-H bond angle of $\Phi_{HNH} = 117 \pm 5°$ for the FT-LIED field-dressed NH₃⁺ structure.

Our measured near-planar structure may be due to one or a combination of the following reasons: (i) the field strength, $F$, used in the calculations of Ref. [34] corresponded to a significantly different peak pulse intensity (5.0 x 10¹³ W/cm²), compared to the one used in our experimental conditions (1.3 x 10¹⁴ W/cm²); (ii) the NWP was propagated on field-free potentials[57], neglecting the important effects of the strong laser field; (iii) the model of Ref. [34] also neglects dynamics induced by the strong laser field that may occur in the neutral molecule prior to ionization. To account for the non-negligible contribution of the MIR laser field, we calculated field-dressed Born-Oppenheimer curves with the field strength set to the corresponding peak intensity 1.3 x 10¹⁴ W/cm² (*i.e.* 3.1 V/Å), as shown in Fig. 5. Orange (blue) dashed curves show the field-dressing when the polarization vector is parallel, F > 0 (antiparallel, F < 0) to the static dipole moment of the ammonia molecule. Importantly, the field-free planar equilibrium cationic structure has now been shifted towards a bent field-dressed structure (*i.e.* Q = −14°; $\Phi_{HNH} = 114°$) caused by the strong laser field, dressing the molecule. There is an excellent agreement between our measured FT-LIED field-dressed NH₃⁺ structure ($\Phi_{HNH} = 117 \pm 5°$) and the calculated equilibrium geometry of the field-dressed ground cationic state ($\Phi_{HNH}^{\widetilde{X}\,^2A_2''} = 114°$). It should be noted that the calculations presented in this work are static in nature, and that quantum dynamical calculations will be required to further investigate the time-resolved nature of this field-dressed system, which are planned in future investigations.

## IV. CONCLUSIONS

In summary, we directly retrieve the geometric structure of $NH_3^+$ with picometer spatial and femtosecond temporal resolution using MIR FT-LIED. We use strong external fields (*i.e.* 3.1 V/Å) to investigate the response of an isolated ammonia molecule to strong-field ionization and the subsequently induced pyramidal-to-planar transition dynamics. We identified a near-planar ammonia cation with a H-N-H bond angle of $\Phi_{HNH} = 117 \pm 5°$. We calculate the field-dressed PECs of $NH_3^+$ and show that the equilibrium field-dressed structure is distorted by the intense laser field, compared to the corresponding field-free case ($\Phi_{HNH}^{FF} = 120°$). The minimum of the FD PEC displaced towards a more bent, near-planar structure ($\Phi_{HNH}^{FD} = 114°$), which has a excellent agreement with our FT-LIED measured $NH_3^+$ structure. Additionally, it would be beneficial to study the dynamics of the ammonia system through quantum-dynamical wave-packet calculations that also include the interaction of the molecule with the intense laser field to further confirm the experimental results.

## SUPPLEMENTARY MATERIAL

The supplementary material gives a description of the structural retrieval process and the determination of the uncertainty in the extracted structural parameters.

## ACKNOWLEDGEMENTS

J.B. and group acknowledge financial support from the European Research Council for ERC Advanced Grant "TRANSFORMER" (788218), ERC Proof of Concept Grant "miniX" (840010), FET-OPEN "PETACom" (829153), FET-OPEN "OPTOlogic" (899794), Laserlab-Europe (EU-H2020 654148), MINECO for Plan Nacional FIS2017-89536-P; AGAUR for 2017 SGR 1639, MINECO for "Severo Ochoa" (SEV- 2015-0522), Fundació Cellex Barcelona, CERCA Programme / Generalitat de Catalunya, and the Alexander von Humboldt Foundation for the Friedrich Wilhelm Bessel Prize. J.B., K.A. and R.Moszynski. acknowledge the Polish National Science Center within the project Symfonia, 2016/20/W/ST4/00314. J.B and B.B. acknowledge Severo Ochoa" (SEV- 2015-0522). J.B. and A.S. acknowledge funding from the Marie Sklodowska-Curie grant agreement No. 641272. C.D.L is supported in part by Chemical Sciences, Geosciences and Biosciences Division, Office of Basic Energy Sciences, Office of Science, U. S. Department of Energy under Grant No. DE-FG02-86ER13491. J.S. and S.G. highly acknowledges support from the European Research Council (ERC) for the ERC Consolidator Grant QUEM-CHEM (772676). The authors thank Alejandro Saenz for helpful discussions.

## DATA AVAILABILITY

The data that support the findings of this study are available from the corresponding author upon reasonable request.

## REFERENCES

1.  Checover, S. *et al.* Dynamics of the proton transfer reaction on the cytoplasmic surface of bacteriorhodopsin. *Biochemistry* **40**, 4281–4292 (2001).
2.  Parks, S. K., Chiche, J. & Pouysségur, J. Disrupting proton dynamics and energy metabolism for cancer therapy. *Nat. Rev. Cancer* **13**, 611–623 (2013).

3. Xu, J., Blaga, C. I., Agostini, P. & DiMauro, L. F. Time-resolved molecular imaging. *J. Phys. B At. Mol. Opt. Phys.* **49**, 112001 (2016).
4. Ischenko, A. A., Weber, P. M. & Miller, R. J. D. Capturing Chemistry in Action with Electrons: Realization of Atomically Resolved Reaction Dynamics. *Chem. Rev.* **117**, 11066–11124 (2017).
5. Amini, K. & Biegert, J. Ultrafast electron diffraction imaging of gas-phase molecules. in *Advances in Atomic, Molecular and Optical Physics* **69**, 163–231 (Academic Press Inc., 2020).
6. Oberhammer, H. *The Electron Diffraction Technique, Part A von: Stereochemical Applications of Gas-Phase Electron Diffraction. Berichte der Bunsengesellschaft für physikalische Chemie* **93**, (John Wiley & Sons, Ltd, 1989).
7. Ladd, M. & Palmer, R. *Structure determination by X-ray crystallography: Analysis by X-rays and neutrons. Structure Determination by X-ray Crystallography: Analysis by X-rays and Neutrons* (Springer US, 2013).
8. Derome, A. E. *Modern NMR techniques for chemistry research*. (Elsevier, 1987).
9. Rischel, C. *et al.* Femtosecond time-resolved X-ray diffraction from laser-heated organic films. *Nature* **390**, 490–492 (1997).
10. Minitti, M. P. *et al.* Imaging Molecular Motion: Femtosecond X-Ray Scattering of an Electrocyclic Chemical Reaction. *Phys. Rev. Lett.* **114**, 255501 (2015).
11. Srinivasan, R., Lobastov, V. A., Ruan, C.-Y. & Zewail, A. H. Ultrafast Electron Diffraction (UED). *Helv. Chim. Acta* **86**, 1761–1799 (2003).
12. Otto, M. R., Renée de Cotret, L. P., Stern, M. J. & Siwick, B. J. Solving the jitter problem in microwave compressed ultrafast electron diffraction instruments: Robust sub-50 fs cavity-laser phase stabilization. *Struct. Dyn.* **4**, 051101 (2017).
13. Wolf, T. J. A. *et al.* The photochemical ring-opening of 1,3-cyclohexadiene imaged by ultrafast electron diffraction. *Nat. Chem.* **11**, 504–509 (2019).
14. Ihee, H. *et al.* Direct imaging of transient molecular structures with ultrafast diffraction. *Science* **291**, 458–62 (2001).
15. Yang, J. *et al.* Imaging CF3I conical intersection and photodissociation dynamics with ultrafast electron diffraction. *Science (80-. ).* **361**, 64–67 (2018).
16. Yang, J. *et al.* Simultaneous observation of nuclear and electronic dynamics by ultrafast electron diffraction. *Science (80-. ).* **368**, 885–889 (2020).
17. Centurion, M. Ultrafast imaging of isolated molecules with electron diffraction. *J. Phys. B At. Mol. Opt. Phys.* **49**, 062002 (2016).
18. Shen, X. *et al.* Femtosecond gas-phase mega-electron-volt ultrafast electron diffraction. *Struct. Dyn.* **6**, 054305 (2019).
19. Amini, K. *et al.* Imaging the Renner–Teller effect using laser-induced electron diffraction. *Proc. Natl. Acad. Sci. U. S. A.* **116**, 8173–8177 (2019).
20. Pullen, M. G. *et al.* Influence of orbital symmetry on diffraction imaging with rescattering electron wave packets. *Nat. Commun.* **7**, 11922 (2016).
21. Pullen, M. G. *et al.* Imaging an aligned polyatomic molecule with laser-induced electron diffraction. *Nat. Commun.* **6**, 7262 (2015).
22. Ito, Y. *et al.* Extracting conformational structure information of benzene molecules via laser-induced electron diffraction. *Struct. Dyn.* **3**, 034303 (2016).
23. Liu, X. *et al.* Imaging an isolated water molecule using a single electron wave packet. *J. Chem. Phys.* **151**, 024306 (2019).
24. Wolter, B. *et al.* Ultrafast electron diffraction imaging of bond breaking in di-ionized acetylene. *Science* **354**, 308–312 (2016).
25. Xu, J. *et al.* Diffraction using laser-driven broadband electron wave packets. *Nat. Commun.* **5**, 4635 (2014).
26. Blaga, C. I. *et al.* Imaging ultrafast molecular dynamics with laser-induced electron

diffraction. *Nature* **483**, 194–197 (2012).
27. Lin, C. D., Le, A.-T., Chen, Z., Morishita, T. & Lucchese, R. Strong-field rescattering physics—self-imaging of a molecule by its own electrons. *J. Phys. B At. Mol. Opt. Phys.* **43**, 122001 (2010).
28. Xu, J., Chen, Z., Le, A.-T. & Lin, C. D. Self-imaging of molecules from diffraction spectra by laser-induced rescattering electrons. *Phys. Rev. A* **82**, 033403 (2010).
29. Meckel, M. *et al.* Laser-induced electron tunneling and diffraction. *Science* **320**, 1478–82 (2008).
30. Okunishi, M., Niikura, H., Lucchese, R. R., Morishita, T. & Ueda, K. Extracting Electron-Ion Differential Scattering Cross Sections for Partially Aligned Molecules by Laser-Induced Rescattering Photoelectron Spectroscopy. *Phys. Rev. Lett.* **106**, 063001 (2011).
31. Piancastelli, M. N., Cauletti, C. & Adam, M. -Y. Angle-resolved photoelectron spectroscopic study of the outer- and inner-valence shells of $NH_3$ in the 20–80 eV photon energy range. *J. Chem. Phys.* **87**, 1982–1986 (1987).
32. Banna, M. S. & Shirley, D. A. Molecular photoelectron spectroscopy at 132.3 eV. The second-row hydrides. *J. Chem. Phys.* **63**, 4759–4766 (1975).
33. Brion, C. E., Hamnett, A., Wight, G. R. & Van der Wiel, M. J. Branching ratios and partial oscillator strengths for the photoionization of NH3 in the 15–50 eV region. *J. Electron Spectros. Relat. Phenomena* **12**, 323–334 (1977).
34. Kraus, P. M. & Wörner, H. J. Attosecond nuclear dynamics in the ammonia cation: Relation between high-harmonic and photoelectron spectroscopies. *ChemPhysChem* **14**, 1445–1450 (2013).
35. Sayres, S. G., Ross, M. W. & Castleman, A. W. Influence of clustering and molecular orbital shapes on the ionization enhancement in ammonia. *Phys. Chem. Chem. Phys.* **13**, 12231–12239 (2011).
36. Förster, J. & Saenz, A. Theoretical Study of the Inversion Motion of the Ammonia Cation with Subfemtosecond Resolution for High- Harmonic Spectroscopy. *ChemPhysChem* **14**, 1438–1444 (2013).
37. Förster, J., Plésiat, E., Magaña, Á. & Saenz, A. Imaging of the umbrella motion and tunneling in ammonia molecules by strong-field ionization. *Phys. Rev. A* **94**, 043405 (2016).
38. Woywod, C., Scharfe, S., Krawczyk, R., Domcke, W. & Köppel, H. Theoretical investigation of Jahn-Teller and pseudo-Jahn-Teller interactions in the ammonia cation. *J. Chem. Phys.* **118**, 5880–5893 (2003).
39. Belyaev, A. K., Domcke, W., Lasser, C. & Trigila, G. Nonadiabatic nuclear dynamics of the ammonia cation studied by surface hopping classical trajectory calculations. *J. Chem. Phys.* **142**, 104307 (2015).
40. Viel, A., Eisfeld, W., Neumann, S., Domcke, W. & Manthe, U. Photoionization-induced dynamics of ammonia: Ab initio potential energy surfaces and time-dependent wave packet calculations for the ammonia cation. *J. Chem. Phys.* **124**, 214306 (2006).
41. P, D. CRC Handbook of Chemistry and Physics. *J. Mol. Struct.* **268**, 320 (1992).
42. Thai, A., Hemmer, M., Bates, P. K., Chalus, O. & Biegert, J. Sub-250-mrad, passively carrier–envelope-phase-stable mid-infrared OPCPA source at high repetition rate. *Opt. Lett.* **36**, 3918 (2011).
43. Colosimo, P. *et al.* Scaling strong-field interactions towards the classical limit. *Nat. Phys.* **4**, 386–389 (2008).
44. Keldysh, L. V. Ionization in the field of a strong electromagnetic wave. *J. Exptl. Theor. Phys.* **20**, 1945–1957 (1965).
45. Ullrich, J. *et al.* Recoil-ion and electron momentum spectroscopy : *Reports Prog.*

*Phys.* **66**, 1463–1545 (2003).
46. Wolter, B. *et al.* Strong-Field Physics with Mid-IR Fields. *Phys. Rev. X* **5**, 021034 (2015).
47. Purvis, G. D. & Bartlett, R. J. A full coupled-cluster singles and doubles model: The inclusion of disconnected triples. *J. Chem. Phys.* **76**, 1910–1918 (1982).
48. Y. Shao, Z. Gan, E. Epifanovsky, A. T. B. Gilbert, M. Wormit, J. Kussmann, A. W. Lange, A. Behn, J. D. *et al.* Advances in molecular quantum chemistry contained in the Q-Chem 4 program package. *Mol. Phys.* **113**, 184–215 (2015).
49. Dunning, T. H. Gaussian basis sets for use in correlated molecular calculations. I. The atoms boron through neon and hydrogen. *J. Chem. Phys.* **90**, 1007–1023 (1989).
50. Roos, B. O., Lindh, R., Malmqvist, P. Å., Veryazov, V. & Widmark, P. O. Main Group Atoms and Dimers Studied with a New Relativistic ANO Basis Set. *J. Phys. Chem. A* **108**, 2851–2858 (2004).
51. F. Aquilante, J. Autschbach, R. K. Carlson, L. F. Chibotaru, M. G. Delcey, L. De Vico, I. Fdez. Galván, N. Ferré, L. M. Frutos, L. Gagliardi, M. Garavelli, A. Giussani, C. E. Hoyer, G. Li Manni, H. Lischka, D. Ma, P. Å. Malmqvist, T. Müller, A. Nenov, M., R. L. MOLCAS 8: New Capabilities for Multiconfigurational Quantum Chemical Calculations across the Periodic Table. *J. Comput. Chem.* **37**, 506–541 (2016).
52. Kaiser, J. & Schafer, R. On the use of the I0-sinh window for spectrum analysis. *IEEE Trans. Acoust.* **28**, 105–107 (1980).
53. Smith, J. O. (Julius O. *Mathematics of the discrete Fourier transform (DFT) : with audio applications*. (W3K Publishing, 2007).
54. Nelson, D. D., Fraser, G. T. & Klemperer, W. Ammonia dimer: A surprising structure. *J. Chem. Phys.* **83**, 6201–6208 (1985).
55. Beu, T. A. & Buck, U. Vibrational spectra of ammonia clusters from n=3 to 18. *J. Chem. Phys.* **114**, 7853–7858 (2001).
56. Kuchitsu, K., Guillory, J. P. & Bartell, L. S. Electron-Diffraction Study of Ammonia and Deuteroammonia. *J. Chem. Phys.* **49**, 2488–2493 (1968).
57. Kraemer, W. P. & Špirko, V. Potential energy function and rotation-vibration energy levels of NH3+. *J. Mol. Spectrosc.* **153**, 276–284 (1992).

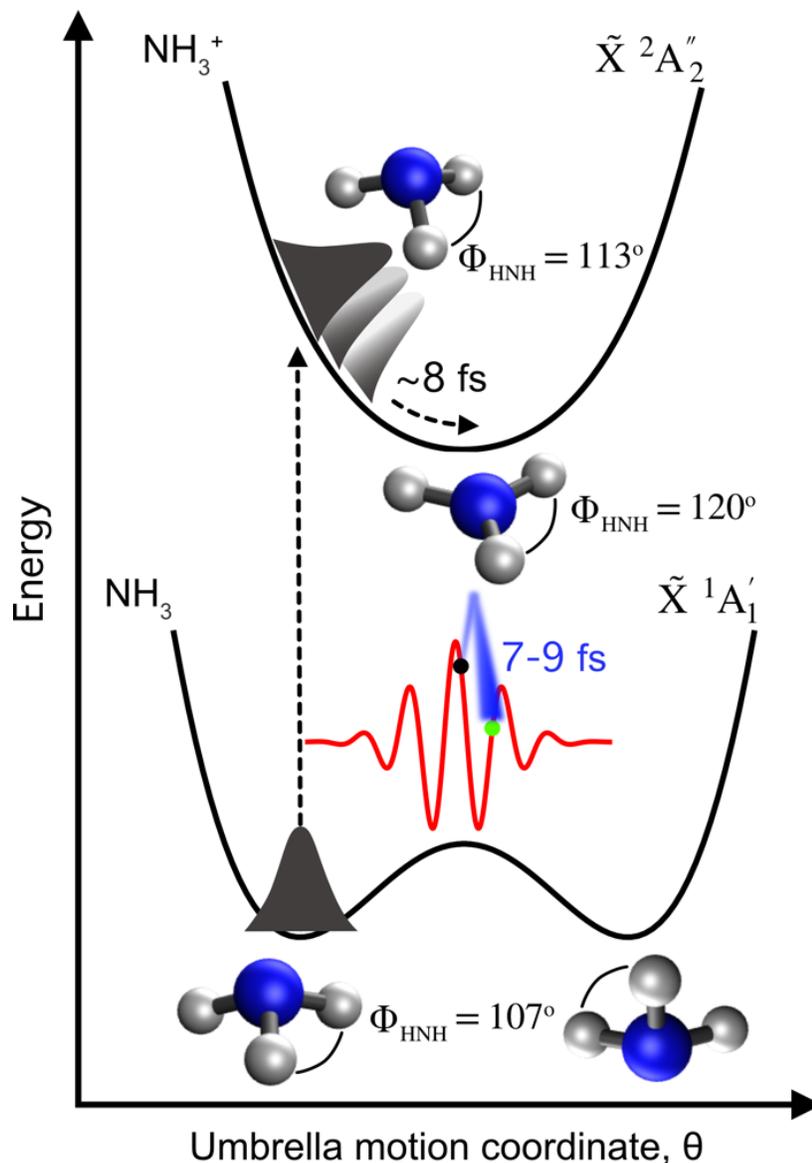

**Fig. 1 | Scheme of the ultrafast umbrella motion of ammonia.** The potential energy curves for the neutral (cation) NH₃ (NH₃⁺) in the ground electronic state $\tilde{X}\,^1A'_1$ ($\tilde{X}\,^2A''_2$) are shown. Upon strong-field ionization, a nuclear wave packet is launched into the electronic ground state of the cation, $\tilde{X}\,^2A''_2$, reaching the minimum of the cation's potential energy curve on a predicted ~8-fs timescale. We probe the geometric structure of the NH₃⁺ cation by ionizing a neutral NH₃ molecule (see black dot) and emitting an electron wave packet (blue shaded) that recollides back onto the target NH₃⁺ ion 7.8 – 9.8 fs after ionization (see green dot) through the MIR-LIED process. The HNH bond angle, $\Phi_{HNH}$, for the structures shown is indicated.

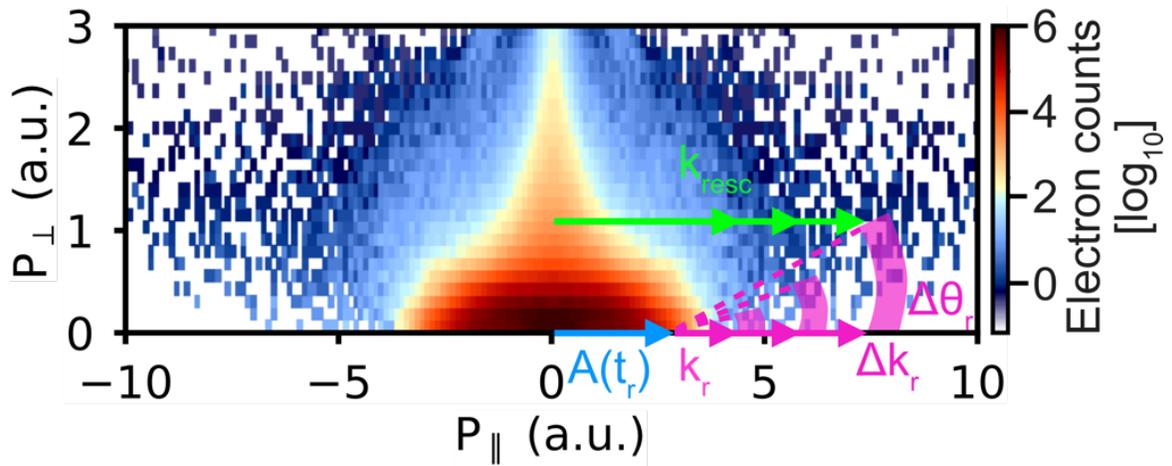

**Fig. 2 | FT-LIED extraction.** Logarithmically scaled momentum distribution of electrons detected in coincidence with $NH_3^+$ fragments only, given in atomic units (a.u.). The return momentum, $k_r$, at the time of rescattering, $t_r$, is obtained by subtracting the vector potential, $A(t_r)$, from the detected rescattering momentum, $k_{resc}$, given by $k_r = k_{resc} - A(t_r)$. The energy-dependent interference signal is extracted by integrating the area indicated by a block arc at different vector potential kicks along $P_\perp = 0$. The block arc is given by a small range of rescattering angles and momenta, $\Delta\theta_r$ and $\Delta k_r$, respectively. We used $\Delta k_r = 0.2$ a.u. together with progressively increasing $\Delta\theta_r$ values from 2º to 10º with increasing $k_{resc}$.

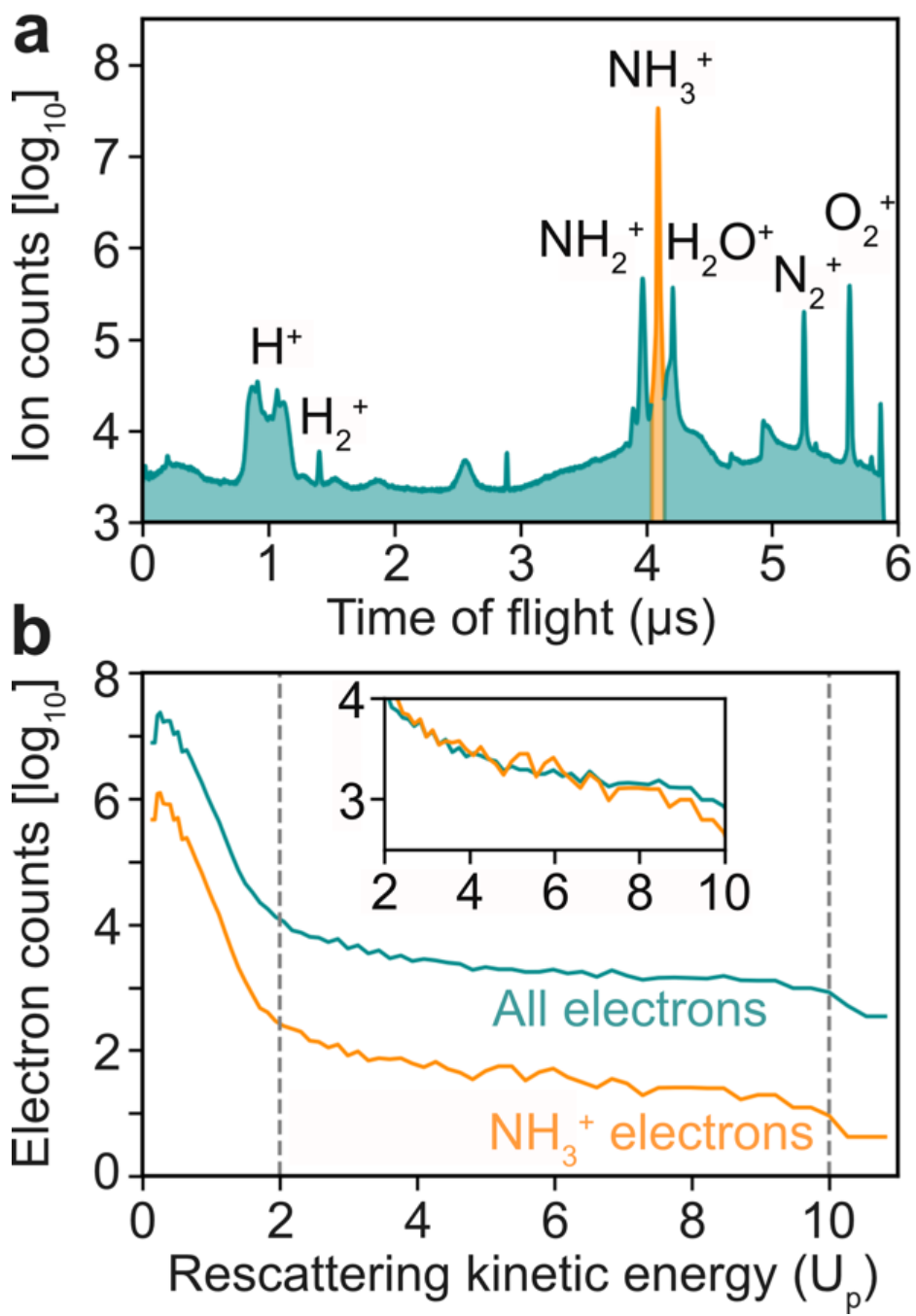

**Fig. 3 | Electron-ion coincidence detection. (a)** Ion time-of-flight (ToF) spectrum, with the main ToF peak near 4.1 μs corresponding to the molecular ion of interest, $NH_3^+$ (orange). **(b)** The signal as a function of rescattered kinetic energy given in ponderomotive energy, $U_p$, for all electrons (petrol blue) and those electrons detected in coincidence with $NH_3^+$ (orange). The inset shows a detailed view of the electron signal in the rescattering regime (*i.e.* 2–10 $U_p$) with both distributions overlaid on top of each other, highlighting the importance of coincidence detection; the $NH_3^+$ electron distribution was scaled by a factor of 50.

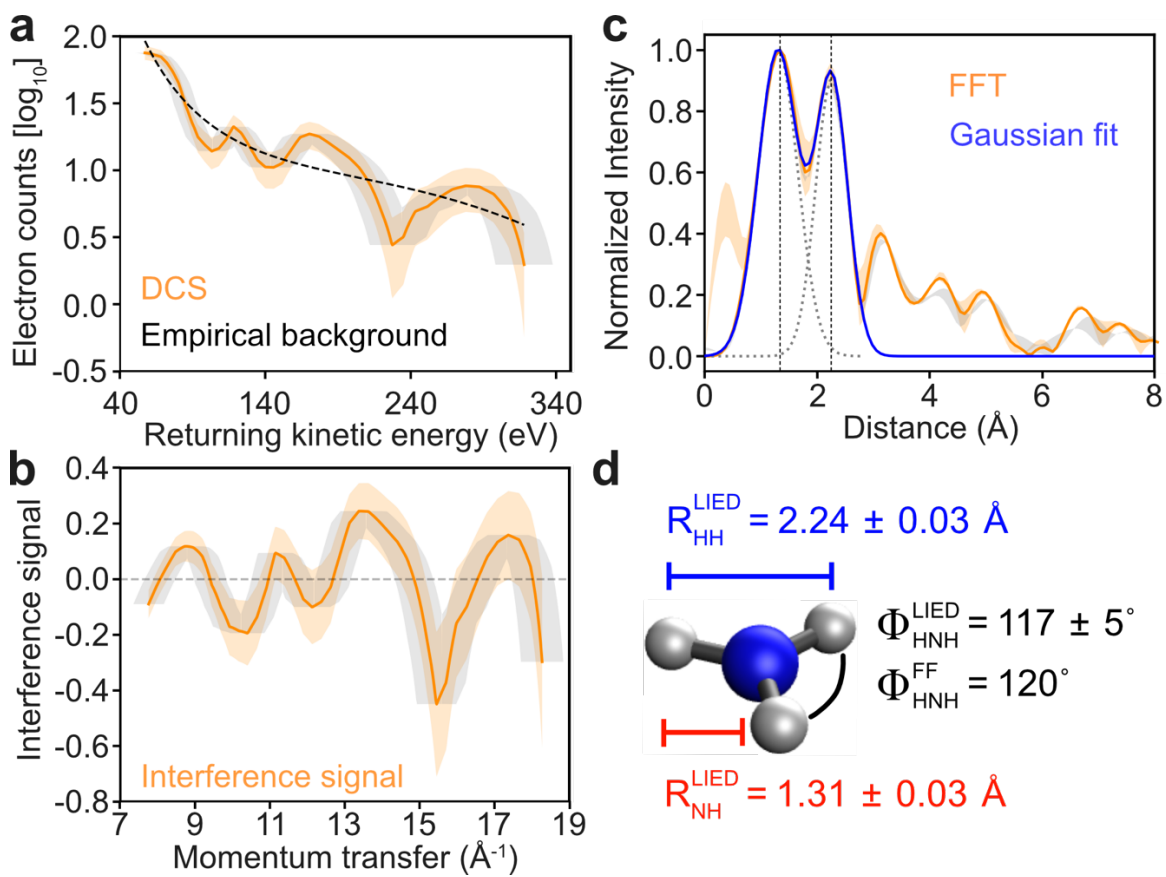

**Fig. 4 | Molecular structure retrieval. (a)** Modulated total interference signal (orange solid trace) and its estimated Poissonian statistical error (orange shaded region) together with the measured longitudinal momentum error (gray shaded region) are shown with the background atomic signal empirically extracted using a third-order polynomial fit (black dotted trace). **(b)** LIED interference signal plotted as a function of momentum transfer, $q = 2k_\mathrm{r}$, in the back-rescattered frame. The orange shaded region corresponds to the Poissonian statistical error and the gray shaded region to the detected longitudinal momentum error. **(c)** Fast Fourier spectrum (blue solid trace) along with the individual (gray dashed traces) and sum (blue solid trace) of Gaussian fits. Orange (gray) shaded region represents the FFT spectra for the two extrema of the Poissonian (momentum) error. The black vertical dotted lines indicate the mean centre positions of the two Fourier peaks corresponding to the N-H and H-H internuclear distances, $R_\mathrm{NH}$ and $R_\mathrm{HH}$, respectively. We note that the peaks above 3 Å could arise from the formation of clusters and are in accordance with Refs.[54] and [55] . **(d)** Sketch of measured NH$_3^+$ LIED structure together with field-free equilibrium NH$_3^+$ ground state structure is shown. The following geometrical parameters were extracted: $R_\mathrm{NH} = 1.31 \pm 0.03$ Å; $R_\mathrm{HH} = 2.24 \pm 0.03$ Å; and H-N-H bond angle, $\Phi_\mathrm{HNH} = 117 \pm 5°$.

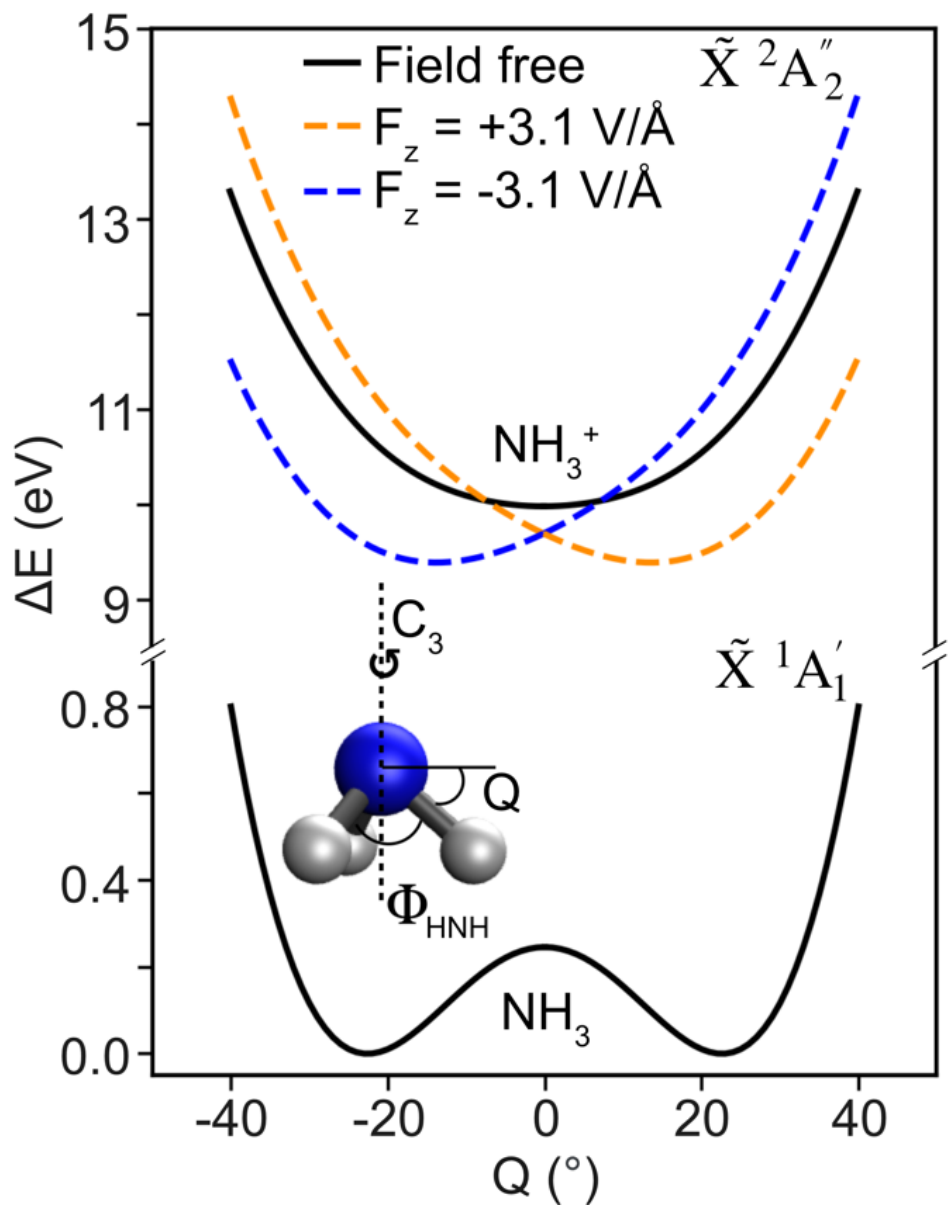

**Fig. 5 | Quantum chemistry calculations.** Ground state field-free PECs along the inversion coordinate for neutral NH$_3$ ($\tilde{X}^1A'_1$) (bottom, solid black curve) and cation NH$_3^+$ ($\tilde{X}^2A''_2$) (top, solid black curve), and the corresponding field-dressed curves (dashed colored traces) with a field strength of 3.1 V/Å (*i.e.* $I_0 = 1.3 \times 10^{14}$ W/cm$^2$). The polarization vector points along the static dipole moment of the molecule. Field-dressed PECs for the field pointing parallel to the dipole moment vector of the molecule (orange dashed curve; F>0) and antiparallel (blue dashed curve; F<0) are shown. On the bottom left, a sketch of the geometry of the ammonia molecule, showing the H-N-H bond angle, $\Phi_{HNH}$, the inversion coordinate employed, $Q$, and the $C_3$ rotation axis.

# Supplementary Material: Laser-induced electron diffraction of the ultrafast umbrella motion in ammonia

B. Belsa[1], K. Amini[1,2], X. Liu[1], A. Sanchez[1] T. Steinle[1], J. Steinmetzer[3], A.T. Le[4], R. Moshammer[5], T. Pfeifer[5], J. Ullrich[5], R. Moszynski[2], C.D. Lin[6], S. Gräfe[3], J. Biegert[1,7,†]

[1]*ICFO - Institut de Ciencies Fotoniques, The Barcelona Institute of Science and Technology, 08860 Castelldefels (Barcelona), Spain.*
[2]*Department of Chemistry, University of Warsaw, 02-093 Warsaw, Poland.*
[3]*Institute of Physical Chemistry and Abbe Center of Photonics, Friedrich-Schiller-Universität Jena, Helmholtzweg 4, 07743 Jena, Germany.*
[4]*Department of Physics, Missouri University of Science and Technology, Rolla, MO 65409.*
[5]*Max-Planck-Institut für Kernphysik, Saupfercheckweg 1, 69117, Heidelberg, Germany.*
[6]*Department of Physics, J. R. Macdonald Laboratory, Kansas State University, 66506-2604 Manhattan, KS, USA.*
[7]*ICREA, Pg. Lluís Companys 23, 08010 Barcelona, Spain.*

[†]To whom correspondence should be addressed to. Email: jens.biegert@icfo.eu.

**STRUCTURAL RETRIEVAL**

The Fast Fourier transform (FFT) spectrum shown in Fig. 4c is generated for a range of different $E_{\min}$–$E_{\max}$ windows in the experimental DCS, spanning $E_{\min} = 1.5 - 2.4\,U_p$ and $E_{\max} = 8.0 - 10.0\,U_p$. We generate a $10 \times 10$ grid of the variation in the retrieved N-H bond length, $R_{\text{NH}}$, and the H-H internuclear distance, $R_{\text{HH}}$, for the aforementioned range of $E_{\min}$ and $E_{\max}$. This generates 100 FFT spectra each of which contain two Fourier peaks. The center position of the two Fourier peaks is extracted by fitting a sum of two gaussian distributions, and the center position of the gaussian fits give the corresponding $R_{\text{NH}}$ and $R_{\text{HH}}$ values. Thus, 100 values for each $R_{\text{NH}}$ and $R_{\text{HH}}$ are generated for the 100 different $E_{\min}$–$E_{\max}$ windows. A weighted average of the X-Y internuclear $R_{\text{XY}}$ is calculated as given by

$$\bar{R}_{\text{XY}} = \frac{\sum_{i=1}^{N} w_i\, R_{\text{XY},i}}{\sum_{i=1}^{N} w_i} \quad [1],$$

where $N$ is the number of $E_{\min}$–$E_{\max}$ windows (i.e. 100), and $w_i$ is the calculated weight given by

$$w_i = \frac{1}{\sigma_{G,i}^2} \quad [2],$$

which is inversely proportional to the square of the gaussian fit error, $\sigma_{G,i}$. This error corresponds to the variance of the parameter estimate in the gaussian fit to our data and it is given by the square root of the main diagonal elements obtained in the covariance matrix.

The variance of the weighted average can be calculated using propagation of errors as

$$\sigma_G^2 = \frac{1}{\sum_{i=1}^{N}\frac{1}{\sigma_{G,i}^2}} \quad [3].$$

The deviation of the of the 100 $R_{XY}$ values from its mean $\bar{R}_{XY}$ value is calculated as its standard deviation, $\sigma_{SD}$, given as

$$\sigma_{SD} = \sqrt{\frac{\sum_{i=1}^{N}(R_{XY,i} - \bar{R}_{XY})^2}{N-1}} \quad [4].$$

The errors in $R_{XY}$ from the standard deviation and gaussian fit are propagated using

$$\sigma_{XY} = \sqrt{\sigma_{SD}^2 + \sigma_G^2} \quad [5].$$

Thus, the mean values of the N-H and H-H internuclear distances ($\bar{R}_{NH}$ and $\bar{R}_{HH}$) are obtained together with their corresponding errors ($\sigma_{NH}$ and $\sigma_{HH}$).

This procedure is then repeated four more times in which a DCS is used that corresponds to the maximum and minimum extrema of the Poisson statistical error (orange shaded area in Fig. 4a) and longitudinal momentum error (gray shaded area in Fig. 4a) DCS distributions given by Eq. [6] and Eq. [7], respectively

$$\delta z = d[\log(DCS)] \approx 0.434 \frac{\delta DCS}{DCS} \quad [6],$$

$$\delta q = 2\delta k_r \quad [7].$$

As such, five values of mean internuclear distance and corresponding errors are calculated for the experimental DCS and the maximum and minimum extrema of the Poisson and momentum error distributions. The simple average of the five $\bar{R}_{XY}$ means is taken as the final reported $R_{XY}$ value (see Fig. 4d) together with the corresponding final error which is given by

$$\text{error} = \frac{\sqrt{\sum_{i=1}^{N}\sigma_{XY,i}^2}}{5} \quad [8].$$

The H-N-H bond angle, $\phi_{HNH}$, was determined by geometrical transformation of the averaged $R_{NH}$ and $R_{HH}$, as given by Eq. [9]. Since only two peaks are detected in the FFT spectrum, we assumed that ammonia is symmetrically bent upon the pyramidal-to-planar transition (*i.e.* all three H-N-H bond angles are identical).

$$\phi_{HNH} = 2\sin^{-1}\left(\frac{R_{HH}}{2R_{NH}}\right) \quad [9]$$

The uncertainty in $\phi_{HNH}$, $\delta\phi_{HNH}$, is therefore, given by error propagation as follows in Eq. [10],

$$\delta\phi_{HNH} = \sqrt{\left(\frac{\partial\phi}{\partial R_{HH}}\right)^2 \delta R_{HH}^2 + \left(\frac{\partial\phi}{\partial R_{NH}}\right)^2 \delta R_{NH}^2} \quad [10].$$